\documentclass[journal]{IEEEtran}
\usepackage{amsmath,amsfonts}
\usepackage{algorithmic}
\usepackage{algorithm}
\usepackage{array}
\usepackage[caption=false,font=normalsize,labelfont=sf,textfont=sf]{subfig}
\usepackage{textcomp}
\usepackage{stfloats}
\usepackage{url}
\usepackage{verbatim}
\usepackage{graphicx}
\usepackage{cite}
\usepackage{xcolor}
\usepackage{multirow}
\usepackage{float}

\begin{document}
\title{
A Telecentric Offset Reflective Imaging System (TORIS) for Terahertz Imaging and Spectroscopy
}

\author{Pouyan Rezapoor\(^1\), Aleksi Tamminen\(^1\), Juha Ala-Laurinaho\(^1\), Dan Ruan\(^2\), Zachary Taylor\(^1\) \\
    {\small 
    \(^1\)Aalto University Department of Electronics and Nanoengineering, MilliLab, Espoo, Finland\\
    \(^2\)Department of Radiation Oncology, University of California, Los Angeles, California, USA
    }
}



\maketitle

\begin{abstract}

Terahertz (THz) imaging has emerged as a promising technology in medical diagnostics, thanks to non-ionizing radiation and the high sensitivity of THz waves to water content. However, traditional THz imaging systems face challenges such as slow mechanical scanning, limited field-of-view, and poor telecentricity. To address these limitations, we present the telecentric offset reflective imaging system, a novel dual-mirror scanning design optimized for high-speed, distortion-free imaging. Utilizing a telecentric $f-\theta$ lens and ray-tracing and physical optics simulations, the system achieves uniform resolution across a 50\(\times\)50 mm\(^2\) field of view without mechanical translation stages. The system demonstrates its capability through broadband spectral imaging of a USAF resolution test target across WR-2.2 (325 - 500 GHz) and WR-1.5 (500 - 700 GHz) rectangular waveguide frequency bands, achieving consistent beam focus and minimal distortion, with maximum deviation of 
2.7\(^{\circ}\) from normal incidence and beam waist of 2.1\(\lambda\) at the edge of the field of view in WR-1.5 frequency band.
Additionally, the system's ability to monitor hydration dynamics is validated by imaging wet tissue paper, illustrating its sensitivity to temporal changes in water content. Further, in vivo imaging of human skin after capsaicin patch application reveals localized hydration variations influenced by biochemical responses and adhesive patch removal. These results confirm the scanner’s potential for real-time hydration sensing and dermatological assessments.
By providing high-resolution, real-time THz imaging, TORIS establishes a new standard in THz-based spectroscopy and imaging, with applications spanning clinical diagnostics, wound assessment, and material characterization.
\end{abstract}


\section{Introduction}
Terahertz (THz) technology, operating in the frequency range between 0.1 and 10 THz, has become a promising tool across various fields, including material characterization, security screening, and biomedical imaging. This frequency range, which lies between the microwave and infrared regions, is particularly advantageous because it offers non-ionizing radiation, low photon energy, and the capability to penetrate non-conductive materials, making it suitable for biological applications. The interaction of THz waves with water, which is a major component of biological tissues, is especially significant. Water molecules exhibit strong absorption and high permittivity at THz frequencies, creating a high level of contrast that is essential for imaging and detecting changes in tissue hydration\cite{Singh_2021}.

Medical applications of THz imaging leverage these unique interactions with water to monitor tissue properties in a non-invasive manner. THz imaging has been successfully applied in several areas of medical diagnostics, including burn wound imaging \cite{RatSkin_1}, corneal hydration measurement \cite{cornea_1}, skin hydration monitoring \cite{hydration_1}, and the detection of skin and breast cancer \cite{taylor2011thz}. In each case, the ability of THz waves to differentiate water content allows for precise and localized imaging of tissue states. The permittivity of water at THz frequencies is significantly higher than that of other tissue components, making it the primary contributor to the contrast in THz images. Additionally, water is the only tissue constituent that shows considerable dispersion at THz frequencies, further enhancing the sensitivity of these systems to even slight changes in hydration levels. As a result, THz systems can detect minor variations in water content, providing critical insights into physiological changes and disease states \cite{taylor2011thz}.

The current generation of THz imaging systems primarily includes time-domain spectroscopy (THz-TDS) setups and continuous wave (CW) systems \cite{zhang2020continuous}. These configurations often utilize mechanical scanning mechanisms, such as raster scanning, where the sample is moved in a grid-like pattern under a stationary THz beam. While effective, these methods can be time-consuming and impractical for large or in vivo samples, which cannot be easily mounted or translated. An alternative approach is to move the THz imaging equipment itself; however, the size and weight of the components make this a challenging and limited solution.

Efforts have also been made to develop electronic and telecentric beam-steering systems that can scan the beam across the sample surface without physically moving the sample or the equipment \cite{harris2020design}. These systems offer the advantage of maintaining a consistent focal spot size and minimizing phase distortions as the beam scans the target. Common approaches include using telecentric \(f-\theta\) lens designs to scan planar targets effectively, ensuring that the beam remains parallel to the optical axis in the conjugate telecentric plane, and maintaining uniform resolution across the imaging field. Despite these advancements, such systems still face challenges, such as maintaining telecentricity over large fields of view and achieving fast, reliable scanning rates without distortions.

To address these limitations, we propose the Telecentric Offset Reflective Imaging System (TORIS) in this work.
Our method involves a dual-mirror scanning configuration optimized through ray-tracing simulations to minimize distortions and maintain high telecentricity across a broad imaging field. The system is designed to provide a fast scanning speed and consistent beam focus, enabling efficient imaging without the need for mechanical translation stages. By steering the beam electronically using a carefully engineered mirror setup, our design maintains the spot size and ensures that the chief ray remains parallel to the optical axis.

To validate the efficiency of TORIS system, we conducted experiments on two different hydration monitoring scenarios. First, we imaged a wet tissue paper sample as it dried over time, allowing us to track real-time hydration changes. Second, we performed in vivo imaging of human skin following capsaicin patch application to assess localized hydration variations influenced by biochemical responses and adhesive patch removal. This study involving human subjects was approved by the Ethical Review Board of Aalto University. The results confirmed the effectiveness of our THz imaging setup for hydration monitoring, highlighting its capability to capture subtle temporal changes in hydration levels in both biological and non-biological samples.


This work highlights the potential of advanced THz imaging systems for medical diagnostics, emphasizing the importance of water content variation as a contrast mechanism. The TORIS system addresses the current limitations of THz imaging systems and demonstrates its utility in capturing hydration changes with high precision and speed.

In the following sections, we detail the design, implementation, and validation of the proposed THz imaging system. Section II presents the design methodology of the TORIS system, including beam collimation, telecentric lens optimization, and scanning system architecture. Section III describes the experimental setup and imaging procedure, including broadband spectral imaging and hydration sensitivity testing. Section IV discusses the measurement results, highlighting the system’s high-resolution imaging capabilities, hydration sensing performance, and its ability to detect subtle hydration variations in biological tissues, such as capsaicin-induced skin hydration changes.

\section{THz Imaging System Design and Simulation}
\subsection{General Description}
\begin{figure}[t]
    \centering
    \includegraphics[width=\linewidth]{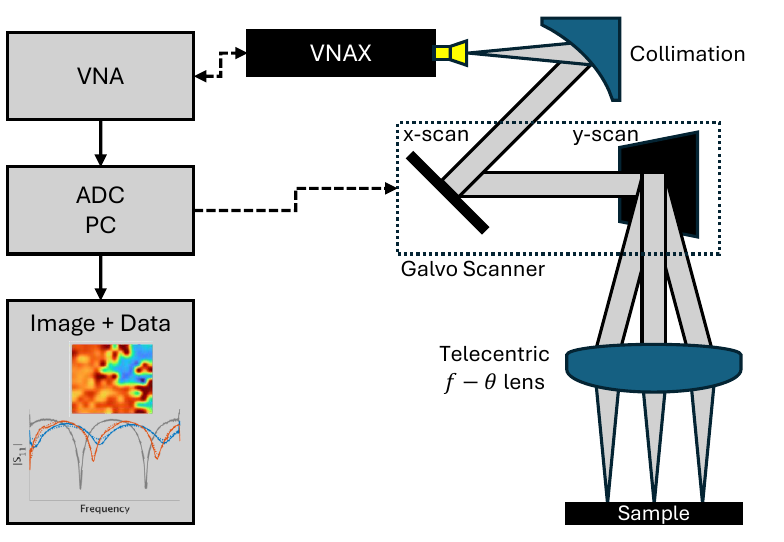}
    \caption{Schematic of the THz imaging system. The system consists of a VNA, VNA Tx/Rx extender, collimator, scanner, and telecentric lens that focuses the THz beam onto the sample.}
    \label{fig1}
\end{figure}

The imaging system is designed to generate, manipulate, and analyze THz waves for high-resolution imaging and spectroscopy. A general schematic of the system is shown in Fig.~\ref{fig1}. The Vector Network Analyzer (VNA) serves as the core signal generator, producing a continuous RF signal that is upconverted to the THz frequency range using VNA extender. The extender also handles the downconversion of reflected THz signals back to the RF domain, allowing the VNA to measure amplitude and phase information for spectral analysis.

Once generated, the THz beam passes through a collimator, which ensures the beam is spatially coherent by transforming it into a parallel Gaussian beam. This collimated beam is then directed into a telecentric lens system that focuses the beam onto the target sample. The telecentric lens ensures normal incidence of the beam on the target plane, which is critical for minimizing aberrations and maintaining a consistent focal spot size. This configuration is particularly advantageous for reflective imaging, as it ensures the reflected beam travels back along its optical path with minimal distortion.

The system incorporates a scanner to achieve precise raster scanning across the target sample along the \textit{x-} and \textit{y-} axes. The reflected beam from the sample is collected by the telecentric lens and collimator and redirected to the VNA extenders for downconversion and measurement.
After each pixel measurement, the VNA sends a trigger pulse to the Analog-to-Digital Converter (ADC), which, in turn, sends a trigger pulse to the scanning system to move to the next position of the scanning.

In the following sections, each of these steps, including the collimation of the Gaussian source beam, the optimization and design of a telecentric lens, and the designing of the scanner geometry are discussed in detail.

\subsection{Beam Collimation}

Broad bandwidth and limited power budget inherent to THz spectroscopic imaging systems make quasi-optical component absorption and dispersion mitigation a high priority. To this end, most systems use off-axis parabolic (OAP) or off-axis elliptical (OAE) mirrors for beam collimation. The most common geometries include OAP with 90\(^{\circ}\) off-set angle, OAP with smaller off-set angle for specific applications, and Dragonian geometry, employing an offset concave hyperboloidal subreflector confocal with an offset paraboloidal main reflector \cite{dragonian_1, dragonian_2}. In this section, after a thorough comparison of the three reflectors, we make a judgment call on which reflector to use based on metrics such as generated cross-polarization, collimated beam Gaussicity, phase variations, and the \(M^2\) parameter \cite{popovic2011thz, rezapoor2022cross ,peiponen2012terahertz, siegman1993defining, siegman1998maybe, friberg1992electromagnetic, tuovinen1992accuracy}.

\subsubsection*{Cross-polarization}
\begin{figure}[t]
    \centering
    \includegraphics[width=\linewidth]{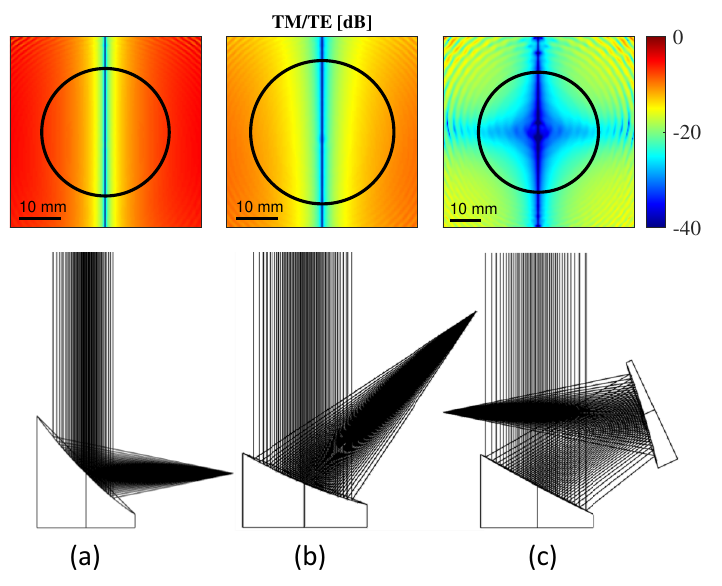}
    \caption{Schematics dual- and single-reflector antennas including (a, b) Off-Axis Parabolic antennas and (c) Dragonian geometry. The ratio of the E-field magnitude in TM polarization over TE polarization inside the collimated beam waist is (a) -8dB, (b) -14dB, and (c) -23dB.}
    \label{fig2}
\end{figure}

Fig.~\ref{fig2} shows the geometry of the Dragonian along with OAP antennas with offset angle of 45\(^{\circ}\) and 90\(^{\circ}\). The E-field magnitude of the collimated beam along transverse electric (TE) and transverse magnetic (TM) polarizations are calculated using physical optics simulations. Here, TE and TM field components are the E-field magnitude along \textit{x}-axis and \textit{z}-axis, respectively. In order to have a measure of the generated cross polarization in the collimated beam, the ratio of the TM field over the TE field is also shown in Fig.~\ref{fig2}. The maximum cross-polarization generated inside the beam waist is -8.5 dB for 90\(^{\circ}\) OAP, -12.28 dB for 45\(^{\circ}\) OAP, and -22.96 dB for the Dragonian geometry.

\subsubsection*{Gaussicity}
The Gaussicity parameter $G$ can also be calculated for the TE polarized field $E(x,y)$ from an overlap integral of the beam and a perfect Gaussian beam $E_G(x,y)$, the waist radius and peak value of which are similar to the beam pattern on the detector plane:

\begin{equation}
G=\frac{\int_{S}^{} |E(x,y)||E_G(x,y)| \,dS}{\sqrt{{\int_{S}^{} |E(x,y)|^2 \,dS}\cdot{\int_{S}^{} |E_G(x,y)|^2 \,dS}}}.
\end{equation}

\begin{table}[b]
    \caption{Beam properties on the detector plane for the Dragonian and OAP reflectors.}
    \label{tab1}
    \centering
    \includegraphics[width=\linewidth]{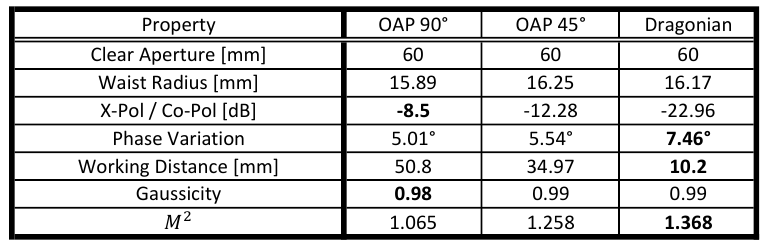}
\end{table}

\subsubsection*{\(M^2\) parameter}
Alongside Gaussicity, the M-squared parameter can also be calculated \cite{siegman1998maybe}. For this purpose, we need to evaluate the second moment or the variance of the beam intensity profile $I(x,y)$ across $x$-axis ($\sigma_x$), which obeys:

\begin{equation}
\sigma_x^2=\frac{\int_{-\infty}^{\infty}(x-x_0)^2I(x,y) \,dxdy}{\int_{-\infty}^{\infty}I(x,y) \,dxdy}
\end{equation}

where $x_0$ is the center of gravity of the beam. By simulating the beam profile at different near-field and far-field distances and calculating the second moment of the beam intensity profile at each position, we can calculate the M-squared parameter by fitting to the following model:

\begin{equation}
W_x^2(z)=W_{0x}^2+M_x^4\times(\frac{\lambda}{\pi W_{0x}})^2(z-z_{0x})^2
\end{equation}

where $W_x$ is the beam spot size, which is twice the variance for a Gaussian beam, i.e. $W_x=2\sigma_x$, $M_x^2$ is the parameter characteristic of the beam, $W_{0x}$ is the beam waist size, $\lambda$ is the wavelength, and $z-z_{0x}$ is the distance from the beam waist. Note that the value of $M_x^2$ for any arbitrary beam is $\geq 1$, and the limit of $M_x^2\equiv1$ occurs only for single-mode $TEM_{0}$ Gaussian beams.

The beam properties are calculated and summarized in Table \ref{tab1} for the three reflector antennas. All the reflectors result in a similar beam waist radius of approximately \(16 \, \mathrm{mm}\) on the detector plane. The Dragonian reflector antenna generated significantly lower cross-polarization but exhibits higher phase variations, a smaller working distance (the distance between the source plane and the aperture plane), and a significantly larger \(M^2\) parameter. In contrast, the \(90^{\circ}\) OAP generates excessive cross-polarization and demonstrates poorer Gaussicity.

The \(45^{\circ}\) OAP is selected as the reflector of choice in this work, as it provides an acceptable balance of cross-polarization, \(M^2\) parameter, phase variations, Gaussicity, and working distance, making it suitable for the intended application.

\subsection{Telecentric \(f-\theta\) Lens Design}

\begin{figure}[b]
    \centering
    \includegraphics[width=\linewidth]{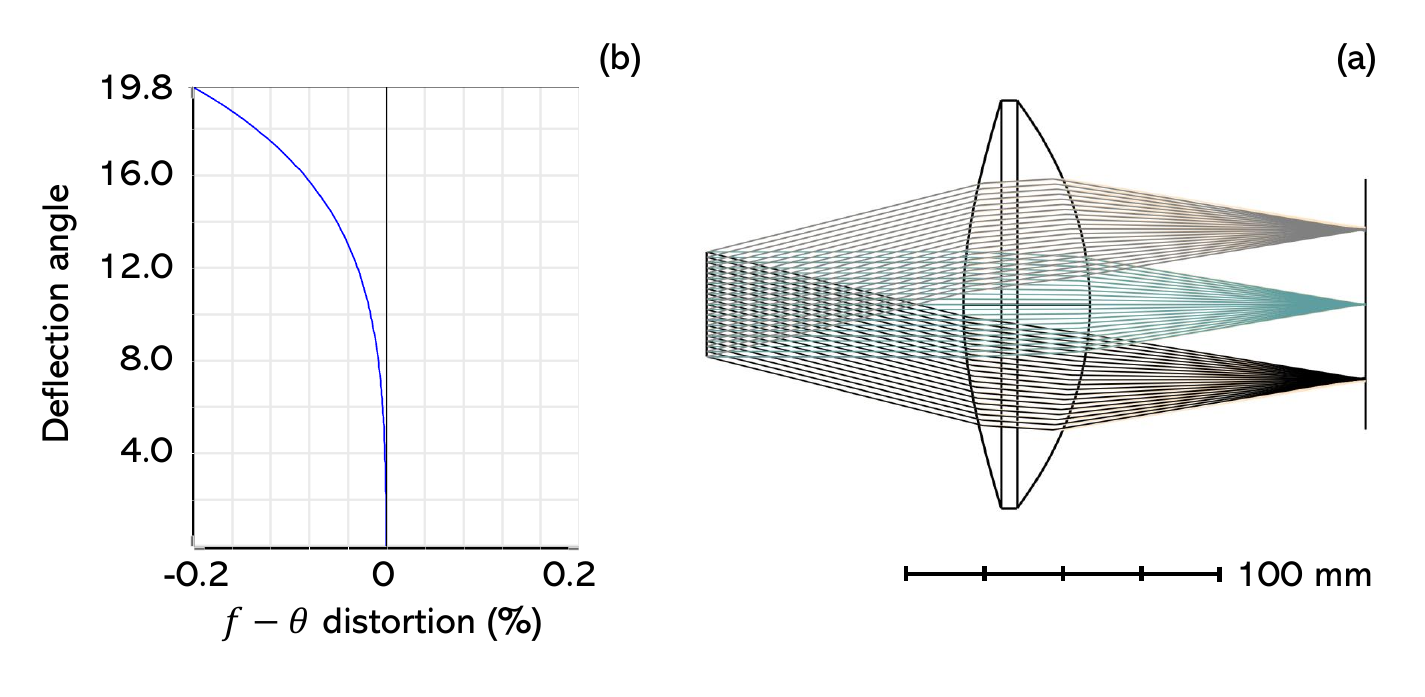}
    \caption{Ray-tracing simulation of the telecentric $f-\theta$ lens design. The figure shows the collimated Gaussian beam being focused onto the target plane. The optimized lens geometry ensures minimal distortion and consistent focal spot size across the field of view.}
    \label{fig3}
\end{figure}

\(f-\theta\) lenses have diverse applications in beam scanning systems, including THz imaging systems where it is necessary to scan and image a planar target plane \cite{rezapoor2022telecentric, harris2020design}. The beam displacement in \(f-\theta\) lenses is a linear function of the deflection angle, resulting in a constant scan rate on the target plane. The focal position on the target plane has a transverse displacement of \(r = f \times \theta\), where \(f\) is the effective focal length and \(\theta\) is the beam's deflection angle.

Additionally, for the beam to reflect from the target plane and travel back to the source plane, the setup must be telecentric in the image space. This ensures that the chief ray is normally incident on the target plane, minimizing aberrations and ensuring consistent reflectivity.

Designing an optimal telecentric \(f-\theta\) lens requires careful consideration of various parameters, including \(f-\theta\) distortion, beam spot size on the target plane, and telecentricity, based on the optical performance requirements of the imaging system. The lens geometry and aspheric constants for the front and back surfaces are optimized using sequential ray tracing simulations in ZEMAX. This optimization is guided by a merit function that accounts for the desired optical properties and constraints.

Fig.~\ref{fig3} illustrates the geometry of the optimized \(f-\theta\) lens along with the \(f-\theta\) distortion with respect to the deflection angle. The beam is scanned to achieve deflection angles of up to \(\pm 14^\circ\) in both directions. Rexolite is selected as the lens material due to its refractive index of \(1.59\) and relatively low absorption in the THz sub-millimeter wavelengths \cite{lamb1996miscellaneous}. The effective focal length of the lens is \(100 \, \mathrm{mm}\), providing sufficient working distance between the imaging plane and the quasi-optical system without significantly increasing the spot size of the focused beam. The lens thickness is approximately \(40.4 \, \mathrm{mm}\).

\begin{figure}[t]
    \centering
    \includegraphics[width=\linewidth]{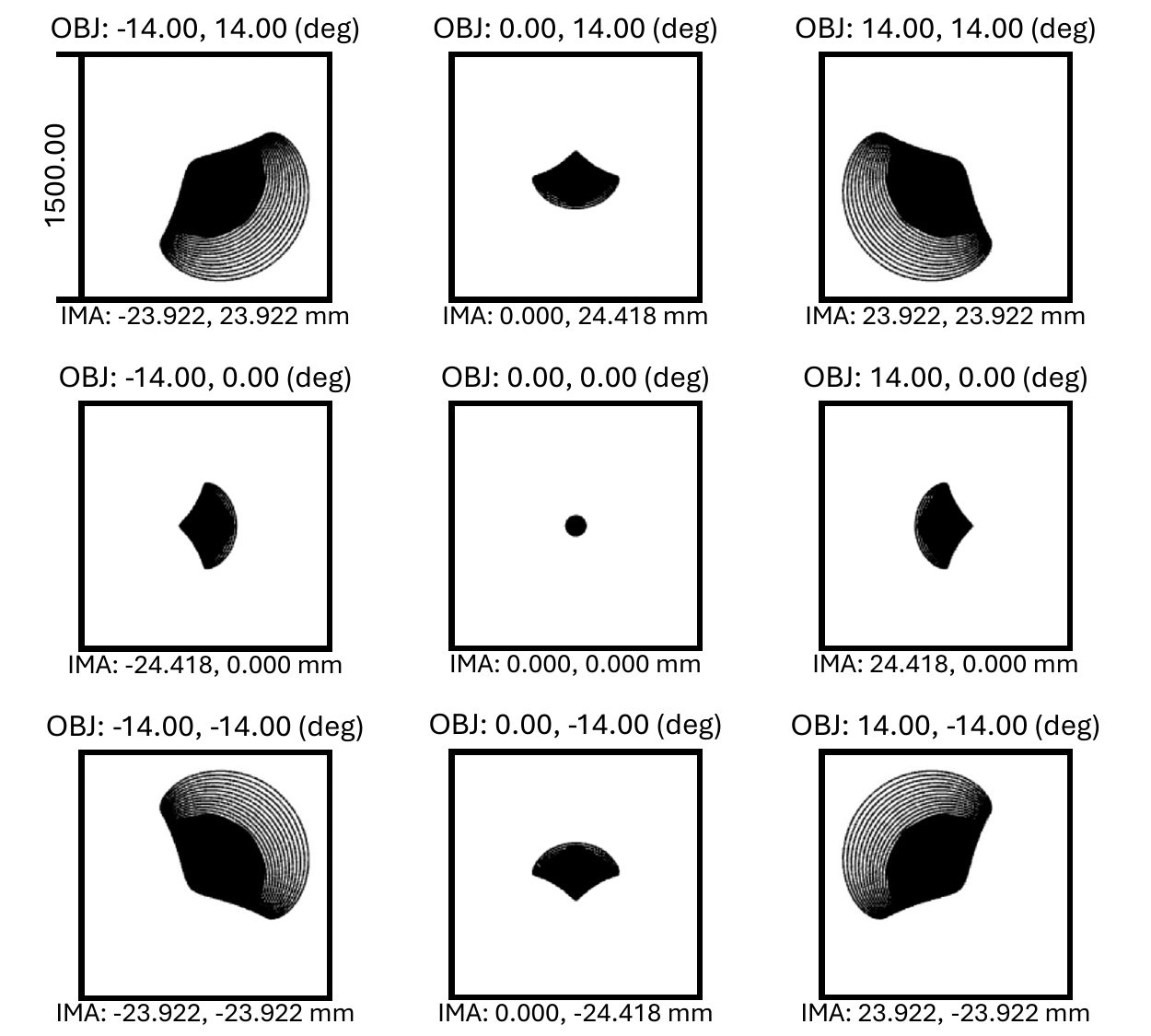}
    \caption{Spot diagram showing the beam profile at the target plane. The root mean square (RMS) spot size is shown at the center and edges of the field of view. The results demonstrate minimal distortion and beam spread, with the smallest spot size at the center of the field.}
    \label{fig4}
\end{figure}

Fig.~\ref{fig3} also highlights that the \(f-\theta\) distortion reaches an acceptable maximum value of approximately \(-0.2\%\) at the edge of the field of view (FOV). The full-field spot diagram (SPD) at the target plane is shown in Fig.~\ref{fig4}. The root mean square (RMS) radius of the focused beam is \(40 \, \mu\mathrm{m}\) at the center and \(147 \, \mu\mathrm{m}\) at the edge of the FOV, demonstrating minimal distortion and consistent beam quality across the imaging area.

\subsection{Scanning System}
\subsubsection*{Common Scanners}

In this section, we utilized the designed telecentric \(f-\theta\) lens to compare three commonly used scanning systems, and integrated the lens with a gimballed mirror scanner and two scanning galvo systems. For each configuration, the geometry was initially optimized using ZEMAX in sequential ray tracing mode to achieve a minimum spot size and optimal telecentricity during beam steering. Subsequently, the scanned beam profile, along with the reflected beam on the source plane, was analyzed using physical optics simulations.

\begin{figure}[t]
    \centering
    \includegraphics[width=\linewidth]{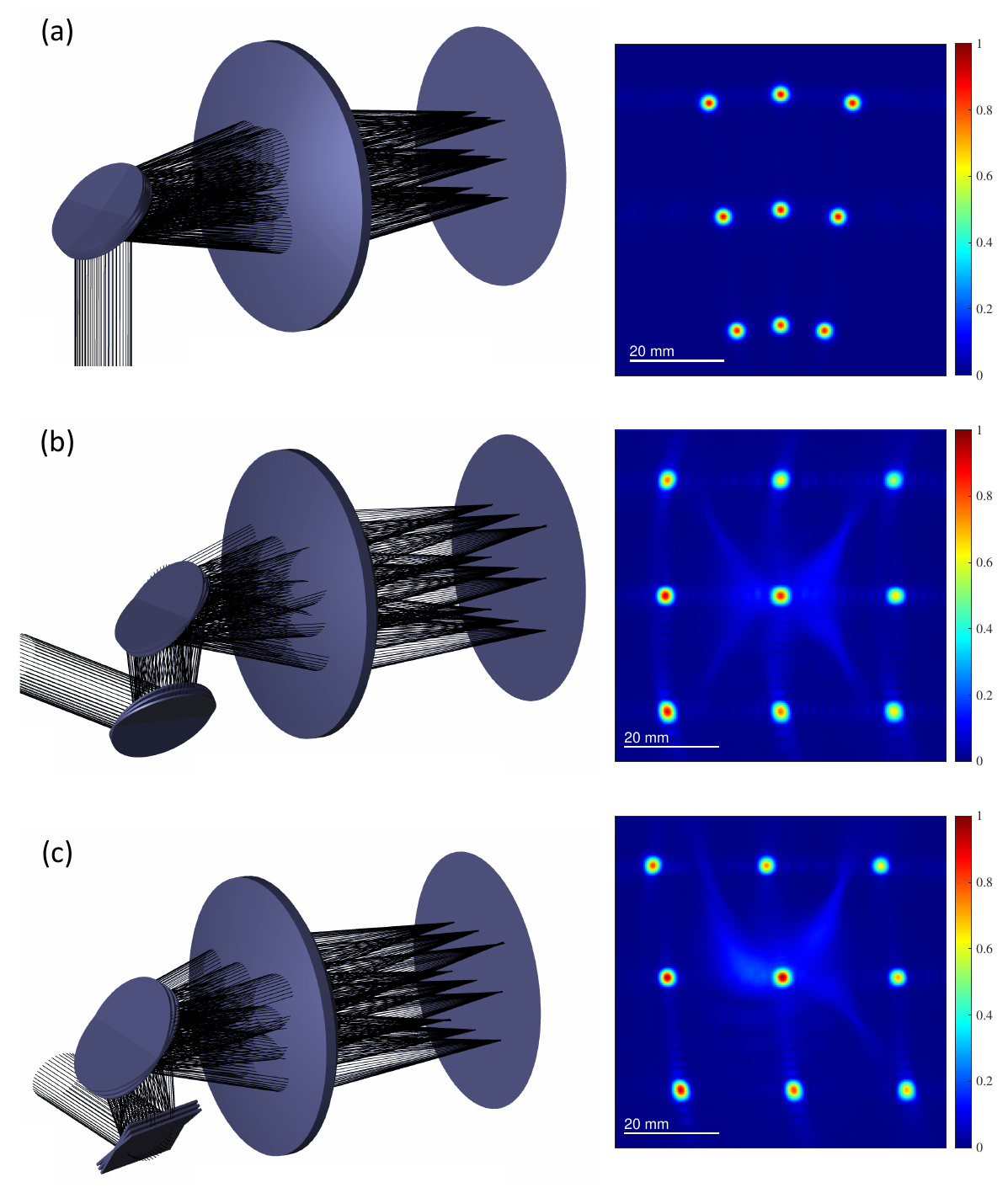}
    \caption{Schematics of three common mirror scanning systems (left side) and the focused beam pattern in physical optics (right side): (a) gimballed mirror scanner, (b) untilted dual-mirror galvo scanner, and (c) tilted dual-mirror galvo scanner. The positioning and scanning geometry of each system are optimized to minimize spot size and maximize telecentricity during beam steering.}
    \label{fig5}
\end{figure}

Schematics of the three scanning systems are shown in Fig.~\ref{fig5}, which include: (a) a gimballed mirror scanner with a single reflector responsible for scanning in both directions, (b) a dual-mirror galvo scanner with two reflectors scanning in one direction, and (c) a tilted dual-mirror galvo scanner, where the first reflector is slightly tilted. The tilted galvo scanner represents a configuration commonly found in commercial scanning systems (e.g., Thorlabs GVS012), and its geometry was included in this study for comparison.

\begin{table}[b]
    \centering
    \caption{Performance of different scanners}
    \includegraphics[width=\linewidth]{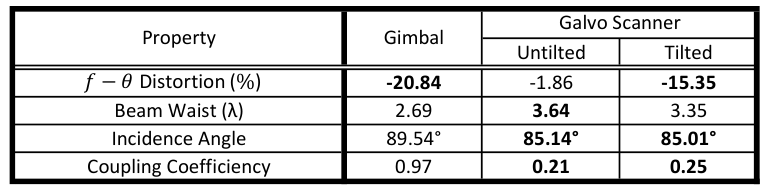}
    \label{table2}
\end{table}

Table~\ref{table2} summarizes the beam properties on the target plane and the coupling efficiency between the initial source beam and the returned beam to the source plane, produced by a reflection from a flat perfectly reflecting mirror. In a gimballed mirror scanning system, the center of the mirror face is positioned at the front focal point of the \(f-\theta\) lens, allowing the lens to focus the beam on the target plane parallel to its optical axis. This results in excellent telecentricity. However, due to the geometry of the gimballed mirror, any elevation change in the beam's direction causes a rotation of the azimuthal axis to a different orientation \cite{harris2020terahertz}. Additionally, the gimballed mirror axes exhibit intercoupling, as reflected in the scanning pattern of the gimballed mirror system shown in Fig.~\ref{fig4}. This intercoupling introduces image distortion, which must be corrected in the imaging system. Moreover, the relatively slow scanning speed of the gimballed mirror system limits its applicability for imaging dynamic phenomena. To overcome this limitation, a dual-mirror scanning system is typically preferred.

In galvo scanner systems, however, only one of the scanning mirrors can be placed in the focal point of the \(f-\theta\) lens, which will create distortions in the focal spot and the chief ray incidence angle will not be perfectly parallel to the lens optical axis, which will reflect in the poor coupling coefficiency of the system at the edge of the FOV, larger \(f-\theta\) distortion, and a poor telecentricity.

\subsubsection*{Proposed TORIS scanner}
\begin{figure}[b]
    \centering
    \includegraphics[width=0.8\linewidth]{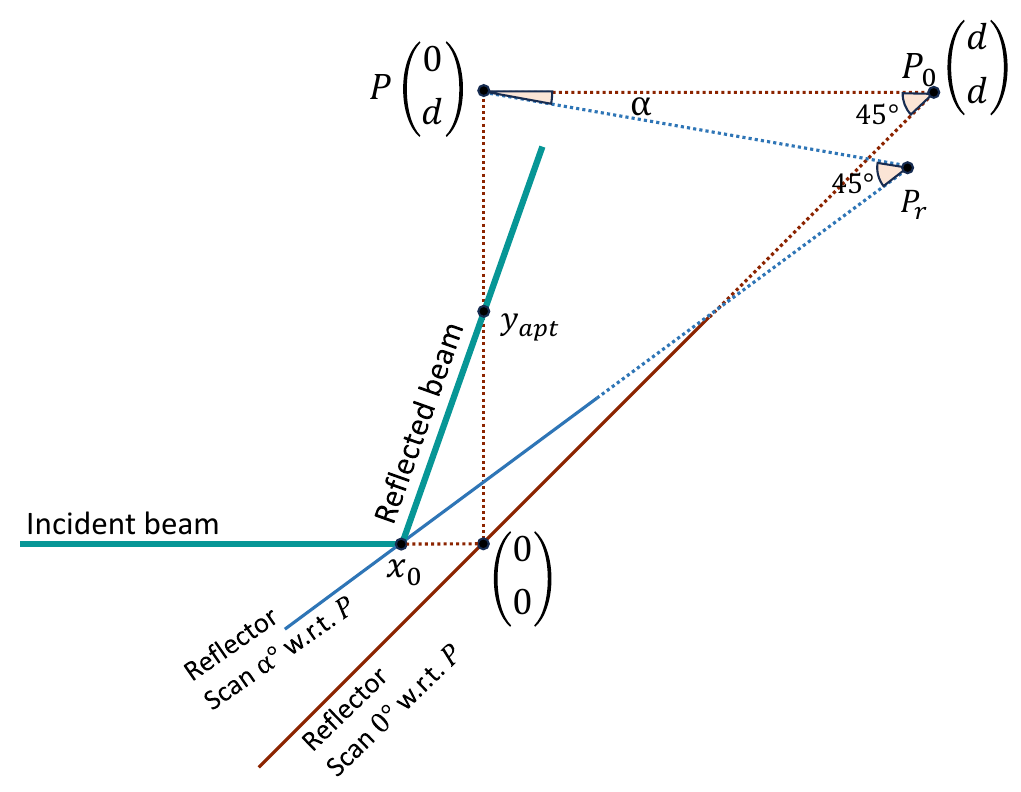}
    \caption{Schematic of the scanned beam geometry when the first reflector is scanned with respect to a pivot point located at distance \(d\) from its center axis. The offset pivot point's effect on aperture plane location and scanned beam path length is illustrated.}
    \label{fig7}
\end{figure}

In a standard galvo scanner system, the scanning axes are positioned in separate aperture planes, making it impossible for both planes to align at the focal point of the telecentric lens. As a result, the design only approximates telecentricity. The degree of deviation from ideal performance depends on the system's intermediate optical configuration and is typically significant only at low zoom settings, where large scan angles are required. This is reflected in the relatively low coupling coefficiency of the scanner, as can be seen in Table~\ref{table2}.

To mitigate the telecentricity issue, we consider the attempt to enhance the scanning pattern of the first reflector and study how the aperture plane changes if we change its rotational axis and scan it with respect to an offset pivot point rather than its own center axis, as in a typical galvo scanner. Fig.~\ref{fig7} shows the case when the first reflector is scanned with respect to the pivot point \(P\) located at a distance \(d\) offset from the reflector. If the reflector is scanned at \(\alpha\) degrees with respect to point \(P\), the new reflector edge is calculated with the rotation matrix \(R(\alpha)\) as:

\begin{equation}
P_r = R(\alpha) \cdot (P_0 - P) + P
\end{equation}

\begin{equation}
P_r = \begin{bmatrix}
\cos\alpha & -\sin\alpha \\
\sin\alpha & \cos\alpha
\end{bmatrix} \cdot (P_0 - P) + P
\end{equation}

\begin{equation}
P_r = \begin{bmatrix}
d \cdot \cos\alpha\\
d - d \cdot \sin\alpha
\end{bmatrix}.
\end{equation}

By knowing the point \(P_r\) and the angle from \(x-\)axis, the line equation of the scanned reflector can be calculated, through which the point \(x_0\) can be extracted, which is the beam incidence point on the scanned reflector:

\begin{equation}
x_0 = d \cdot \cos\alpha - \frac{d - d \cdot \sin\alpha}{\tan(\pi/4 - \alpha)}.
\end{equation}

By knowing the point \(x_r\) and the angle from \(x-\)axis, the line equation of the reflected beam is calculated, being:

\begin{equation}
y(x) = \tan(\pi/2 - 2\alpha) \cdot (x - x_0)
\label{RefEq}.
\end{equation}

\begin{figure}[b]
    \centering
    \includegraphics[width=\linewidth]{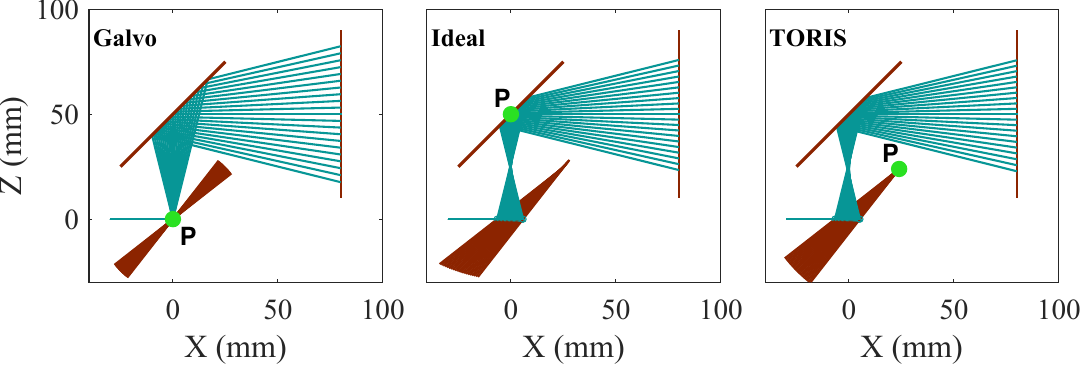}
    \caption{Comparison of chief ray scanning patterns between a typical galvo scanner, ideal pivot point located at the center of the second reflector, and the proposed TORIS scanner.}
    \label{fig6}
\end{figure}

The aperture plane location \(y_{apt}\) is the geometrical point where the reflected beam in the scan center and scan edge are coincident. By inputting \(x = 0\) in (\ref{RefEq}), we have:

\begin{equation}
y_{apt} = \tan(\pi/2 - 2\alpha) \cdot (-x_0) = \frac{-x_0}{\tan(2\alpha)}
\end{equation}

\begin{equation}
=\frac{\frac{d - d \cdot \sin\alpha}{\tan(\pi/4 - \alpha)} - d \cdot \cos\alpha}{\tan(2\alpha)}
\end{equation}

\begin{equation}
=d \cdot \frac{\frac{1 -\sin\alpha}{\tan(\pi/4 - \alpha)} - \cos\alpha}{\tan(2\alpha)}.
\end{equation}

For small scan angles of \(\alpha < 7^{\circ}\):

\begin{equation}
y_{apt} \approx d \cdot \frac{\frac{1 -\alpha}{\frac{1-\alpha}{1+\alpha}} - (1-\frac{\alpha^2}{2})}{2\alpha}
\end{equation}

\begin{equation}
= d \cdot (\frac{1}{2} + \frac{\alpha}{4}) \approx \frac{d}{2}
\label{AptEq}.
\end{equation}

Therefore, if the rotational axis of the reflector is shifted to a pivot point \(P\) located at a distance \(d\) from its center axis, the aperture plane shifts approximately to the middle of its center axis and the pivot point, located at a distance \(\approx d/2\) from its center axis for small scan angles.

\begin{figure}[t]
    \centering
    \includegraphics[width=\linewidth]{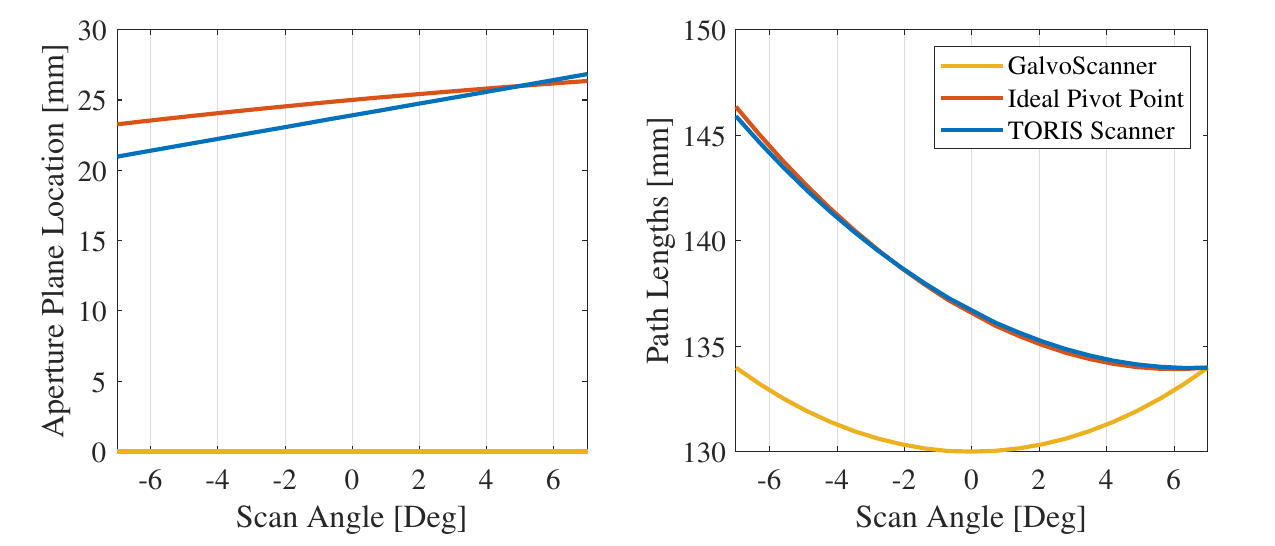}
    \caption{Quantitative comparison of beam path length variations and aperture plane location between the galvo scanner, ideal pivot point system, and the proposed TORIS scanner. The TORIS design shifts the aperture plane towards the second reflector with a cost of larger beam path length.}
    \label{fig8}
\end{figure}

A disadvantage of shifting the rotational axis is that scanned chief rays in the image frame have slightly different optical path lengths. In order to minimize the path length variations, the geometrical optimum point would be to locate the aperture plane of the first reflector axis in the middle of the two scanners.
By concluding this with (\ref{AptEq}), the optimal rotational axis for the first reflector aligns with the rotational axis of the second reflector.
This has the effect of shifting the mirror as it rotates, and improves the accuracy with which the mirrors approach the conjugate telecentric plane, significantly reducing beam aberrations caused by differing focal points in conventional designs.
Fig.~\ref{fig6} illustrates the chief ray scanning patterns for three different scanners including a typical galvo scanner, a scanner with the ideal pivot point for the first mirror, and the TORIS scanner.
It can be seen that the chief ray scanning axis in the two directions are further apart in a typical galvo scanner, which results in the beam aberrations described and discussed previously. By shifting the pivot point to the scanning axis of the second reflector, the chief ray is scanned closer to the second reflector.

However, rotating the first reflector from a point located at the center of another reflector in the scanning system is practically not possible without blocking the collimated beam path. To overcome this issue, we can take a look at the scanning pattern at Fig.~\ref{fig6} and estimate an alternative pivot point which results in a similar scanned beam behavior compared to the ideal pivot point. On this base, we propose the TORIS scanner, where ray tracing optimizations are used to define the pivot point out of the collimated beam path, while resulting in a similar aperture plane, maintaining the telecentricity, \(f-\theta\) distortion, and beam spot size requirements. This offset point is shown in Fig.~\ref{fig6}.

In a typical galvo scanning system, the aperture plane is in the center axis of the first reflector, further increasing the beam aberrations. Locating the rotation axis of the first reflector on the center of the second reflector moves the aperture plane to the middle of the two reflectors, which results in improved accuracy with which the mirrors approach the conjugate telecentric plane. This is illustrated in Fig.~\ref{fig8}, comparing the scanned beam path length and the distance between the two aperture planes in our proposed TORIS scanner with a typical galvo scanner system and the ideal case of having the pivot point in the center of the second reflector. As a summary, the TORIS system has the advantage of improved telecentricity over typical galvo scanning systems, with comparably larger beam path length variations as a trade-off.

\begin{figure}[t]
    \centering
    \includegraphics[width=\linewidth]{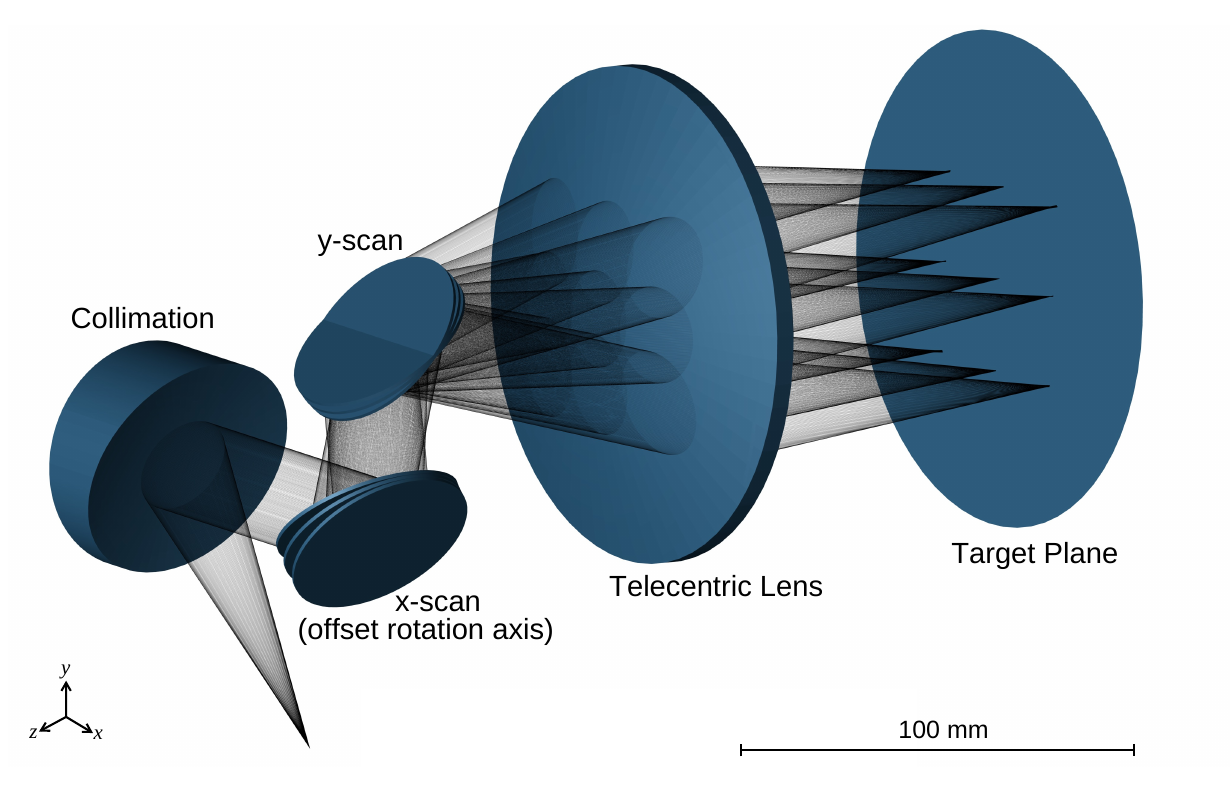}
    \caption{Final geometry of the TORIS quasi-optics optimized in ZEMAX. The collimated beam is steered using a dual-mirror setup, and the focused beam is directed onto the target plane. The system is designed to maintain high telecentricity and minimal distortion across the field of view.}
    \label{fig9}
\end{figure}

The final geometry of the scanning system is illustrated in Fig.~\ref{fig9}. The source beam is collimated with a \(45^{\circ}\) OAP. The modified galvo scanning system deflects the beam in two directions. The \(f-\theta\) lens later focuses the scanned beam with a normal incidence angle on the target plane and similar to previous scanning systems, the reflected beam travels the same optical path back to the original source plane for \(S_{11}\) measurements and calculation of the coupling coefficient.

\begin{figure}[t]
    \centering
    \includegraphics[width=\linewidth]{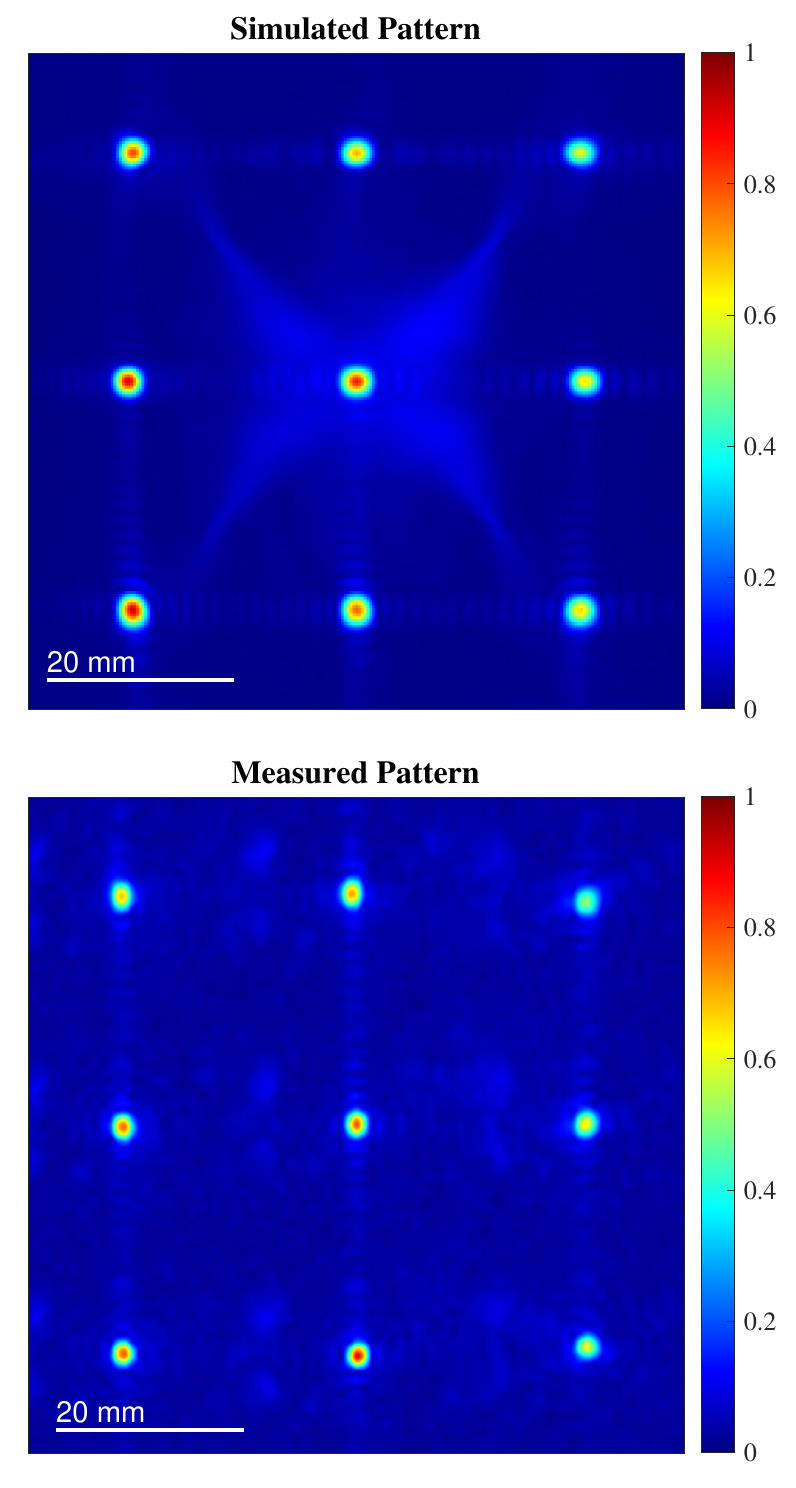}
    \caption{(a) Physical optics simulation and (b) measured scanning pattern on the target plane showing the area covered by the TORIS system at 500 GHz. The beam spot size and telecentricity are optimized to achieve high coupling efficiency and minimal \(f-\theta\) distortion, as calculated in physical optics simulations.}
    \label{fig10}
\end{figure}

The physical optics simulation of the TE field profile of the scanned beam on the target plane is shown in Fig.~\ref{fig10}a, scanning an area of \(52 \times 52\) mm\(^2\).
The beam spot radius on the target plane is \(2.1 \lambda\) at 500 GHz in WR-1.5 band. The coupling coefficient between the source beam and the returned beam which is reflected off of the target plane is \(0.68\) at the edge of the FOV. However, the reflection losses at the lens surfaces and dielectric losses in the lens are not considered in the coupling coefficient calculation. The incidence angle of the chief ray on the target plane at the edge of the FOV is \(87.28^{\circ}\), which is a significant improvement compared to previous galvo scanning systems, reflected also in the higher coupling coefficient. Maximum \(f-\theta\) distortion can also be calculated by considering the chief ray incidence point on the target plane and the deflection angles along each axis. The maximum \(f-\theta\) distortion is \(0.774\%\), a considerably lower value compared to the previous scanners.

\section{Imaging System}
The RF instrumentation is based on an N5225B PNA Microwave Network Analyzer (Keysight Technologies, Inc.). The VNA drives submillimeter wave extenders WR-2.2 and WR-1.5 VNAX (Virginia Diodes Inc.) to achieve the frequency band of 325-500 GHz and 500-700 GHz, respectively. The transceiver VNA extension module (VNAX) was coupled to the Pickett-Potter dual mode horn antenna (Radiometer Physics GmbH) for the corresponding frequency range. The horn antennas serve as a Gaussian radiation source for the OAP collimating mirror (Edmund Optics). The OAP mirror has a diameter of \(D = 50.8 \,\mathrm{mm}\), offset angle of \(\theta = 45^{\circ}\), and apparent focal length of \(f_{p} = 76.2 \,\mathrm{mm}\), which results in an effective focal length of \(f_{e} = 89.27 \,\mathrm{mm}\), calculated from:

\begin{equation}
    f_e = \frac{f_p}{\cos^2{(\frac{\theta}{2})}}
    \label{Eq:OAP}.
\end{equation}

For beam pattern measurements, an open-ended waveguide (OEWG) for the corresponding frequency range was used as a probe antenna coupled to the VNAX receiver module. The VNAX receiver was mounted on a planar near-field scanner (NSI-200 V-5 \(\times\) 5 by Near-Field Systems Inc.) to perform the \(xy\) scanning in the measurement plane. The constructed TORIS scanner uses two hybrid stepper motors for each axis (NEMA 11 frame size from Lin Engineering Inc.) to scan the beam. The measured TE field profile on the target plane is illustrated in Fig.~\ref{fig10}b. The scanner reflectors were machined from aluminum with a polished and fly cut surface to have perfect reflection with minimum surface roughness. The imaging window is a 1-mm thick JGS2 fused silica (University Wafer Inc.) with a refractive index of 1.96 at the operation frequency band \cite{lamb1996miscellaneous}.

\begin{figure}[t]
    \centering
    \includegraphics[width=\linewidth]{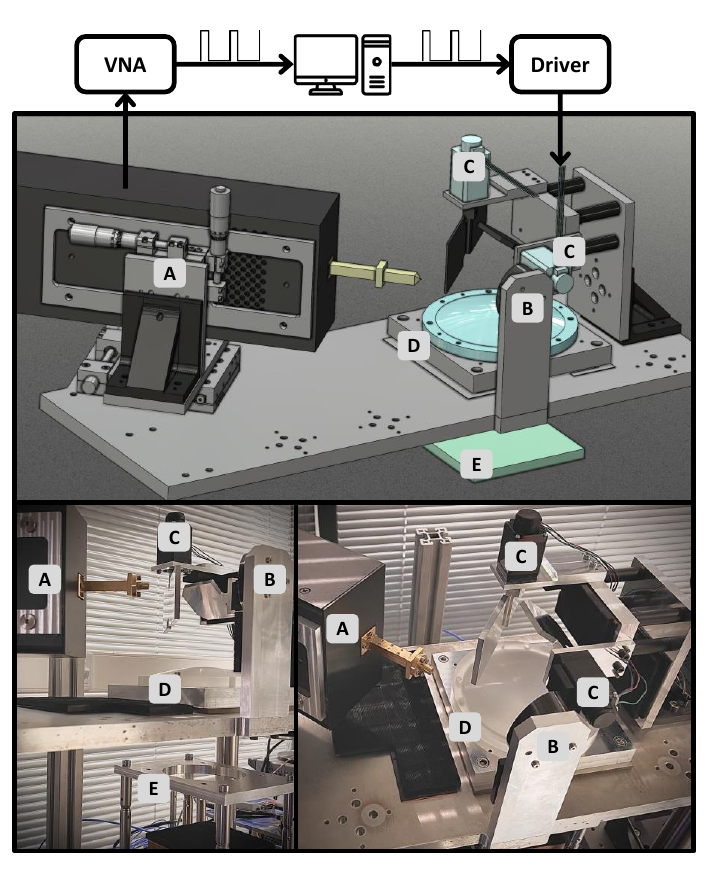}
    \caption{Schematic of the TORIS system using (A) submillimeter wave extender, (B) OAP for beam collimation, (C) scanners mounted on stepper motors, (D) telecentric lens for focusing the beam, and (E) the imaging window.}
    \label{fig11}
\end{figure}

The imaging setup is built on a single base plate, which can be assembled on optical breadboards in two different orientations for lens calibration and imaging. Before the measurements, the VNAX transceiver module with the horn antenna was aligned with the OAP mirror focal point by scanning the field without the galvo scanners present. Once the measurement demonstrated sufficient phase planarity, the VNAX transceiver module was fixed in place. Later, the galvo scanner was attached to the base plate and the base plate was oriented for finding the optimum distance between the lens and the galvo scanner. This distance was then tuned by scanning the near-field pattern at the image side of the lens. Near-field to near-field transformation was used to find the focal plane of the telecentric lens. The lens location was adjusted by inserting 0.5-mm thick shims between the lens support plate and the base plate. Once the measured profile on the target plane achieved the minimum spot size, the lens location and the number of shims inserted were fixed, and the distance between the lens and the imaging plane was also recorded. The base plate was then rotated back to the original state, and the fused silica window was placed on the target plane for imaging. The imaging system is illustrated in Fig.~\ref{fig11}.

The VNA was in Fast-CW mode, using FIFO (first-in first-out) to allow external point-triggering of the stepper motors. The external trigger in the VNA was used to send a TTL (Transistor-Transistor Logic) step pulse to the PC after each sweep. This pulse was then used to trigger the drivers of the stepper motors to move to the next imaging location. After this, when the VNA sweeps the frequency band again, reflection from the new imaging location is recorded. This process is repeated until the whole FOV is imaged. The second reflector which was being scanned from the center of the optical path was responsible for the fast scanning in the \(x\)-axis, while the first reflector,having an offset rotational axis, was responsible for the slow scanning in the \(y\)-axis.

\section{Measurement Results}
\subsection{Test Target Images}

The reflection from the region of interest (RoI) was extracted from the measured \( S_{11} \) by subtracting the mean \( S_{11} \) response in the time domain to improve contrast in the generated images. Time gating was applied and the reflections from the quasi-optical components outside the target location were excluded and only signals within the time window of 3.6 to 4 ns were retained. This time window corresponds to an optical path length range of 55 to 60 cm approximately matching the cumulative path from the source plane to the off-axis parabolic (OAP) mirror 90 mm from the OAP to the scanning mirrors (200 mm), from the scanners to the telecentric lens (150 mm), and from the lens to the imaging window (100 mm).

\begin{figure}[t]
    \centering
    \includegraphics[width=\linewidth]{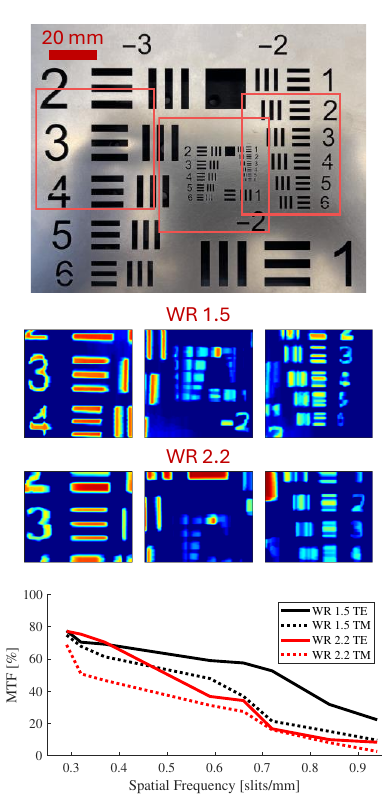}
    \caption{Imaging results of the USAF resolution target in the two rectangular waveguide frequency bands WR-2.2 and WR-1.5.}
    \label{fig12}
\end{figure}

Fig.~\ref{fig12} shows the measured images from the USAF resolution test target with an xy scanning resolution of 1mm. The USAF target is a standard test pattern commonly used to evaluate the resolution and imaging quality of optical systems. The modulation transfer function (MTF) can also be calculated for the images captured at both bands. The small enough difference between the TE and TM polarization suggests acceptably low cross-polarization on the target plane, which mainly arises from choosing \(45^{\circ}\) OAP instead of \(90^{\circ}\) OAP.

The slow scanning of the first reflector from an offset point by a stepper motor, however, causes small high-frequency vibrations to the reflector's movement, which can introduce noise and disturb the generated images. To address this issue, we adopt a directional denoising scheme along the scan axis relative to the slow reflector. Specifically, total variation denoising is applied to one-dimensional (1D) discrete signals with an efficient forward filtering method ~\cite{condat2013direct}. 

\subsection{Total Variation Denoising}

Total Variation (TV) denoising is an important method in signal and image processing  to remove or alleviate oscillatory noise while preserving edges.  TV denoising balances the need to preserve fidelity to the  measurement data and regularize the (piecewise) smoothness of the final denoised signal as characterized by the total variation energy. 

For a one-dimensional (1D) discrete signal \( y = (y[1], \dots, y[N]) \), TV denoising computes the denoised signal \( x^* = (x^*[1], \dots, x^*[N]) \), where \( x^* \) is the solution to the following minimization problem:

\begin{align}
\min_{x \in \mathbb{R}^N} \frac{1}{2} \sum_{k=1}^{N} \left( y[k] - x[k] \right)^2 + \lambda \sum_{k=1}^{N-1} \left| x[k+1] - x[k] \right|,
\end{align}

where \( \lambda \geq 0 \) is a regularization parameter that controls the trade-off between fidelity to the original noisy signal and the smoothness of the output. When \( \lambda \) is small, the denoised signal remains close to the original signal but might retain some noise. As \( \lambda \) increases, the solution becomes smoother, possibly losing some fine details.

\begin{figure}[t]
    \centering
    \includegraphics[width=\linewidth]{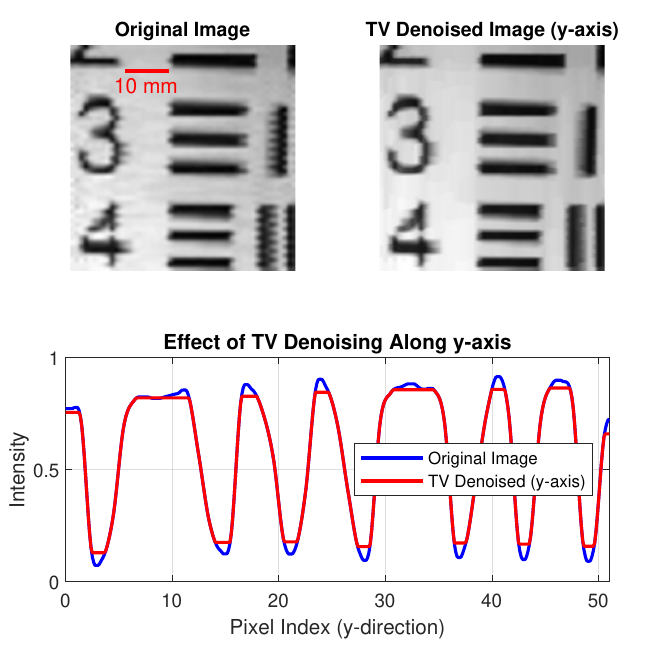}
    \caption{Comparison of the original noisy image and the TV-denoised image (y-axis only). The top row shows the original image (left) and the denoised image (right), demonstrating noise reduction while preserving structural details. The bottom plot illustrates the intensity profile along a selected vertical column, highlighting the smoothing effect of TV denoising without excessive blurring.}
    \label{fig13}
\end{figure}

This TV denoising formulation is a convex optimization problem, ensuring the existence of a unique solution, which adds to its robustness in practical applications. Utilizing the insight from a primal-dual formulation, we adopt a discrete filtering approach for efficient noni-iterative implementation ~\cite{condat2013direct}.

To assess the effectiveness of TV denoising applied along the y-axis, we computed the Peak Signal-to-Noise Ratio (PSNR) and the Edge Preservation Index (EPI). The PSNR of the denoised image was 33.34 dB, indicating a significant reduction in noise while maintaining structural details. Furthermore, the EPI was measured at 0.885, confirming that edge features were well-preserved after denoising. These results demonstrate that the applied TV denoising technique effectively smooths vertical noise fluctuations without introducing excessive blurring or distortion, making it suitable for maintaining fine spatial details in terahertz imaging applications. In Fig.~\ref{fig13}, the difference between the original noisy image and the TV-denoised image can be observed, where the denoising process significantly reduces noise while preserving key image structures.



\subsection{Hydration Sensing}
Hydration assessment using THz imaging is based on the strong absorption and dispersion properties of water in the submillimeter-wave range. To validate the sensitivity of the TORIS system in detecting dynamic hydration changes, we conducted two separate experiments: (1) monitoring the drying process of a wet tissue paper sample and (2) assessing hydration variations in human skin induced by capsaicin application.

The first experiment aimed to demonstrate the system’s capability to track subtle temporal changes in hydration levels by imaging a piece of tissue paper immediately after wetting and over a drying period of several minutes. Since THz waves exhibit high sensitivity to water content, the expected outcome was a gradual increase in reflected intensity as water evaporated, thereby reducing absorption losses.

\begin{figure}[h]
    \centering
    \includegraphics[width=\linewidth]{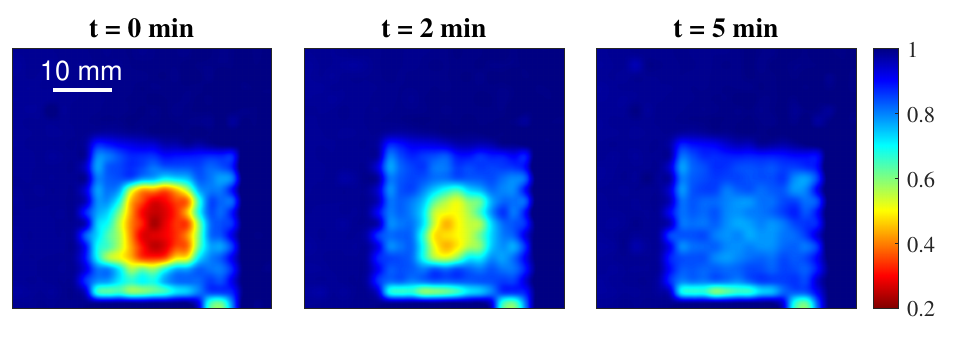}
    \caption{Imaging results of a drying paper sample over several minutes. The system captures changes in reflectivity caused by variations in water content as the paper dries.}
    \label{fig14}
\end{figure}

As shown in Fig.~\ref{fig14}, the acquired THz images at WR-2.2 band confirm a progressive increase in reflectivity over time. The spatial distribution of the reflectivity change is non-uniform, indicating localized differences in evaporation rates. The results underscore the system’s ability to resolve even minor hydration gradients with high sensitivity and spatial precision.

To assess the applicability of the system for hydration sensing in biological tissues, we conducted an in vivo experiment by applying capsaicin patches to the forearm skin. Capsaicin, the active component in chili peppers, is well known for stimulating transient receptor potential vanilloid 1 (TRPV1) channels in sensory nerves, leading to localized vasodilation, increased blood flow, and potential sweat gland activation \cite{capsaicin_1}. Independent of capsaicin’s biochemical effects, applying and removing an adhesive patch also induces skin changes. Mechanical removal of the patch may cause mild epidermal irritation, disrupt the skin barrier, and influence hydration dynamics. These physiological responses contribute to transient changes in skin hydration, which can be probed using THz imaging.

\begin{figure}[h]
    \centering
    \includegraphics[width=\linewidth]{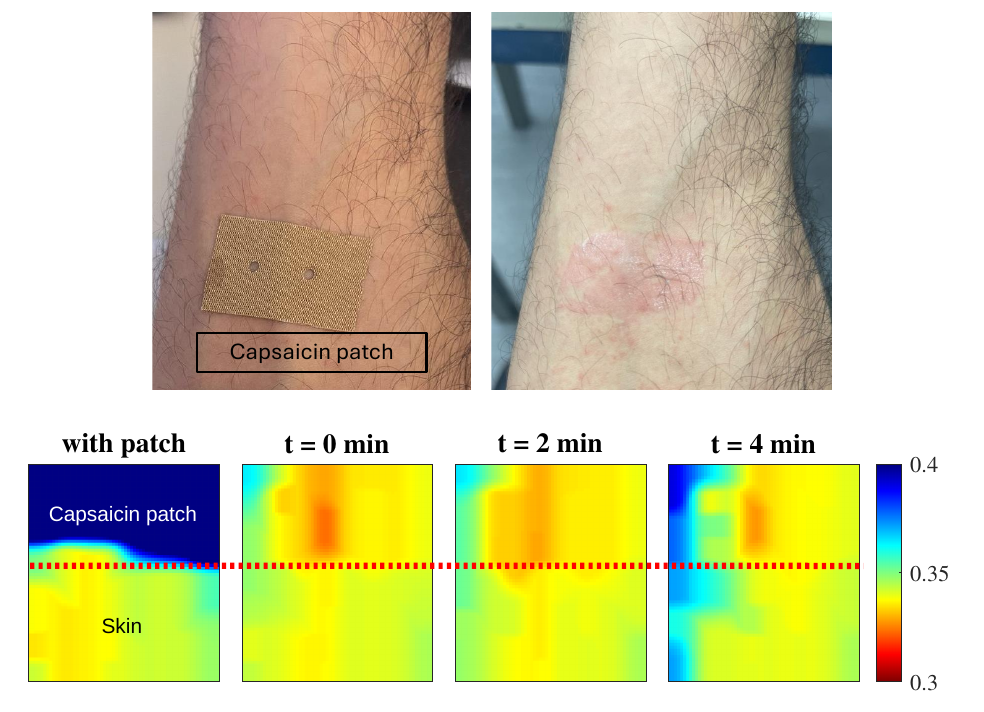}
    \caption{THz imaging results of skin hydration changes following capsaicin patch removal. The reflectivity pattern shows localized variations over time, reflecting capsaicin-induced physiological effects and the mechanical impact of adhesive patch removal on skin hydration. The results confirm the system’s capability to track transient hydration changes in biological tissues.}
    \label{fig15}
\end{figure}

Fig.~\ref{fig15} presents the THz images captured at multiple time intervals after capsaicin patch removal. The reflectivity pattern shows a localized decrease in the applied area, corresponding to changes in skin hydration. The observed trends align with previous studies on THz-based skin hydration sensing, which demonstrate that the hydration levels of the skin increases following application of adhesive patches \cite{patch_emma_1}.

THz reflectivity is a strong indicator of water content variations at sub-millimeter penetration depths. Compared to conventional hydration measurement techniques, such as corneometry or impedance spectroscopy, THz imaging offers the advantage of high spatial resolution and non-invasive, full-field assessment.

\section{Conclusion}

In this work, we introduced the Telecentric Offset Reflective Imaging System (TORIS), a novel dual-mirror scanning configuration designed for terahertz (THz) imaging and spectroscopy. The TORIS system addresses key limitations of conventional THz imaging setups by maintaining consistent telecentricity, achieving high spatial resolution, and providing uniform field-of-view coverage by using two rotators instead of mechanical translation stages. Optimization through ray-tracing and physical optics simulations of the system minimizes distortion and maintains optical efficiency across a broad imaging area. The system demonstrated a maximum deviation of 2.7$^\circ$ from normal incidence, a beam waist of 2.1$\lambda$, and a coupling coefficient of 0.68 at the edge of the field of view in the WR-1.5 frequency band, ensuring minimal distortion, significantly improving coupling efficiency.

To validate the performance of the TORIS system, we conducted a series of experiments. The first set of experiments involved broadband THz spectral imaging of a standard resolution test target, confirming the system's capability to achieve high imaging fidelity across the WR-2.2 (325–500 GHz) and WR-1.5 (500–700 GHz) frequency bands. The second set of experiments focused on hydration sensing at WR-2.2 band, where the system successfully captured temporal hydration changes in two different contexts. First, THz imaging was used to track the drying process of a wet tissue paper sample, demonstrating the system’s sensitivity to dynamic water content variations. Second, an in vivo hydration study was conducted by applying and removing a capsaicin patch from human skin. The THz images revealed localized hydration changes influenced by both the biochemical effects of capsaicin-induced vasodilation and the mechanical disruption of the skin barrier caused by adhesive patch removal.

These results establish TORIS as a highly versatile tool for biomedical imaging, dermatological assessment, wound monitoring, and material characterization, particularly in applications where hydration-sensitive imaging is crucial. The system’s ability to capture transient changes in water content with high spatial resolution and minimal distortion highlights its potential advantages over conventional hydration measurement techniques.

Future work will focus on exploring the spectroscopic features of the TORIS system. Its high-frequency resolution, superior to that of traditional time-domain spectroscopy (TDS) systems, makes it a promising approach for extracting epidermal hydration levels and thickness. By studying the spectroscopic properties of biological tissues, we aim to further advance the system's utility in applications such as skin diagnostics and detailed hydration profiling.

\section*{Conflict of Interest}
None of the authors have a conflict of interest to disclose.

\bibliography{./TORIS_references}

\begin{thebibliography}{10}
\providecommand{\url}[1]{#1}
\csname url@samestyle\endcsname
\providecommand{\newblock}{\relax}
\providecommand{\bibinfo}[2]{#2}
\providecommand{\BIBentrySTDinterwordspacing}{\spaceskip=0pt\relax}
\providecommand{\BIBentryALTinterwordstretchfactor}{4}
\providecommand{\BIBentryALTinterwordspacing}{\spaceskip=\fontdimen2\font plus
\BIBentryALTinterwordstretchfactor\fontdimen3\font minus \fontdimen4\font\relax}
\providecommand{\BIBforeignlanguage}[2]{{%
\expandafter\ifx\csname l@#1\endcsname\relax
\typeout{** WARNING: IEEEtran.bst: No hyphenation pattern has been}%
\typeout{** loaded for the language `#1'. Using the pattern for}%
\typeout{** the default language instead.}%
\else
\language=\csname l@#1\endcsname
\fi
#2}}
\providecommand{\BIBdecl}{\relax}
\BIBdecl

\bibitem{Singh_2021}
I.~Malhotra and G.~Singh, ``Terahertz antenna technology for imaging and sensing applications,'' in \emph{Terahertz Antenna Technology for Imaging and Sensing Applications}.\hskip 1em plus 0.5em minus 0.4em\relax Springer, 2021, pp. 75--102.

\bibitem{RatSkin_1}
P.~Tewari, C.~P. Kealey, D.~B. Bennett, N.~Bajwa, K.~S. Barnett, R.~S. Singh, M.~O. Culjat, A.~Stojadinovic, W.~S. Grundfest, and Z.~D. Taylor, ``In vivo terahertz imaging of rat skin burns,'' \emph{Journal of biomedical optics}, vol.~17, no.~4, pp. 040\,503--040\,503, 2012.

\bibitem{cornea_1}
D.~B. Bennett, Z.~D. Taylor, P.~Tewari, R.~S. Singh, M.~O. Culjat, W.~S. Grundfest, D.~J. Sassoon, R.~D. Johnson, J.-P. Hubschman, and E.~R. Brown, ``Terahertz sensing in corneal tissues,'' \emph{Journal of biomedical optics}, vol.~16, no.~5, pp. 057\,003--057\,003, 2011.

\bibitem{hydration_1}
A.~I. Hernandez-Serrano, X.~Ding, J.~Young, G.~Costa, A.~Dogra, J.~Hardwicke, and E.~Pickwell-MacPherson, ``Terahertz probe for real time in vivo skin hydration evaluation,'' \emph{Advanced Photonics Nexus}, vol.~3, no.~1, pp. 016\,012--016\,012, 2024.

\bibitem{taylor2011thz}
Z.~D. Taylor, R.~S. Singh, D.~B. Bennett, P.~Tewari, C.~P. Kealey, N.~Bajwa, M.~O. Culjat, A.~Stojadinovic, H.~Lee, J.-P. Hubschman \emph{et~al.}, ``Thz medical imaging: in vivo hydration sensing,'' \emph{IEEE transactions on terahertz science and technology}, vol.~1, no.~1, pp. 201--219, 2011.

\bibitem{zhang2020continuous}
Y.~Zhang, C.~Wang, B.~Huai, S.~Wang, Y.~Zhang, D.~Wang, L.~Rong, and Y.~Zheng, ``Continuous-wave thz imaging for biomedical samples,'' \emph{Applied Sciences}, vol.~11, no.~1, p.~71, 2020.

\bibitem{harris2020design}
Z.~B. Harris, S.~Katletz, M.~E. Khani, A.~Virk, and M.~H. Arbab, ``Design and characterization of telecentric f-$\theta$ scanning lenses for broadband terahertz frequency systems,'' \emph{AIP advances}, vol.~10, no.~12, 2020.

\bibitem{dragonian_1}
C.~Dragone, ``Offset multireflector antennas with perfect pattern symmetry and polarization discrimination,'' \emph{Bell System Technical Journal}, vol.~57, no.~7, pp. 2663--2684, 1978.

\bibitem{dragonian_2}
S.~Chang and A.~Prata, ``The design of classical offset dragonian reflector antennas with circular apertures,'' \emph{IEEE Transactions on Antennas and Propagation}, vol.~52, no.~1, pp. 12--19, 2004.

\bibitem{popovic2011thz}
Z.~Popovic and E.~N. Grossman, ``Thz metrology and instrumentation,'' \emph{IEEE Transactions on Terahertz Science and Technology}, vol.~1, no.~1, pp. 133--144, 2011.

\bibitem{rezapoor2022cross}
P.~Rezapoor, A.~Tamminen, J.~Ala-Laurinaho, S.~O. Dabironezare, N.~Llombart, and Z.~Taylor, ``Cross polarization and aberrations with dragonian and equivalent off-axis parabolic mirrors for beam collimation in thz imaging systems,'' in \emph{Passive and Active Millimeter-Wave Imaging XXV}, vol. 12111.\hskip 1em plus 0.5em minus 0.4em\relax SPIE, 2022, pp. 124--130.

\bibitem{peiponen2012terahertz}
K.-E. Peiponen, A.~Zeitler, and M.~Kuwata-Gonokami, \emph{Terahertz spectroscopy and imaging}.\hskip 1em plus 0.5em minus 0.4em\relax Springer, 2012, vol. 171.

\bibitem{siegman1993defining}
A.~E. Siegman, ``Defining, measuring, and optimizing laser beam quality,'' \emph{Laser Resonators and Coherent Optics: Modeling, Technology, and Applications}, vol. 1868, pp. 2--12, 1993.

\bibitem{siegman1998maybe}
------, ``How to (maybe) measure laser beam quality,'' in \emph{Diode Pumped Solid State Lasers: Applications and Issues}.\hskip 1em plus 0.5em minus 0.4em\relax Optica Publishing Group, 1998, p. MQ1.

\bibitem{friberg1992electromagnetic}
A.~T. Friberg, T.~Jaakkola, and J.~Tuovinen, ``Electromagnetic gaussian beam beyond the paraxial regime,'' \emph{IEEE transactions on antennas and propagation}, vol.~40, no.~8, pp. 984--989, 1992.

\bibitem{tuovinen1992accuracy}
J.~Tuovinen, ``Accuracy of a gaussian beam,'' \emph{IEEE transactions on antennas and propagation}, vol.~40, no.~4, pp. 391--398, 1992.

\bibitem{rezapoor2022telecentric}
P.~Rezapoor, A.~Tamminen, J.~Ala-Laurinaho, and Z.~Taylor, ``Telecentric f-theta scanning lens design for terahertz imaging systems,'' in \emph{2022 47th International Conference on Infrared, Millimeter and Terahertz Waves (IRMMW-THz)}.\hskip 1em plus 0.5em minus 0.4em\relax IEEE, 2022, pp. 1--2.

\bibitem{lamb1996miscellaneous}
J.~W. Lamb, ``Miscellaneous data on materials for millimetre and submillimetre optics,'' \emph{International Journal of Infrared and Millimeter Waves}, vol.~17, pp. 1997--2034, 1996.

\bibitem{harris2020terahertz}
Z.~B. Harris, A.~Virk, M.~E. Khani, and M.~H. Arbab, ``Terahertz time-domain spectral imaging using telecentric beam steering and an f-$\theta$ scanning lens: distortion compensation and determination of resolution limits,'' \emph{Optics Express}, vol.~28, no.~18, pp. 26\,612--26\,622, 2020.

\bibitem{condat2013direct}
L.~Condat, ``A direct algorithm for 1-d total variation denoising,'' \emph{IEEE Signal Processing Letters}, vol.~20, no.~11, pp. 1054--1057, 2013.

\bibitem{capsaicin_1}
S.~Munjuluri, D.~A. Wilkerson, G.~Sooch, X.~Chen, F.~A. White, and A.~G. Obukhov, ``Capsaicin and trpv1 channels in the cardiovascular system: the role of inflammation,'' \emph{Cells}, vol.~11, no.~1, p.~18, 2021.

\bibitem{patch_emma_1}
H.~Lindley-Hatcher, J.~Wang, A.~I. Hernandez-Serrano, J.~Hardwicke, G.~Nurumbetov, D.~M. Haddleton, and E.~Pickwell-MacPherson, ``Monitoring the effect of transdermal drug delivery patches on the skin using terahertz sensing,'' \emph{Pharmaceutics}, vol.~13, no.~12, p. 2052, 2021.

\end{thebibliography}
\bibliographystyle{IEEEtran}

\vfill

\end{document}